# Coulomb Drag in Normal Metals and Superconductors: Diagrammatic Approach.

Alex Kamenev and Yuval Oreg

*Department of Condensed Matter, The Weizmann Institute of Science, Rehovot 76100, Israel*

## Abstract

The diagrammatic linear response formalism for the Coulomb drag in two–layer systems is developed. This technique can be used to treat both elastic disorder and intralayer interaction effects. In the absence of intralayer electron–electron correlations we reproduce earlier results, obtained using the kinetic equation and the memory–function formalism. In addition we calculate weak–localization corrections to the drag coefficient and the Hall drag coefficient in a perpendicular magnetic field. As an example of the intralayer interaction effects we consider a situation where one (or both) layers are close to (but above) the superconducting transition temperature. Fluctuation corrections, analogous to the Aslamazov–Larkin corrections, to the drag coefficient are calculated. Although the fluctuation corrections do not enhance the drag coefficient for normal–superconductor systems, a dramatic enhancement is found for superconductor-superconductor structures.

Typeset using REVTEX

## I. INTRODUCTION

In a recent series of experiments the mutual friction between two parallel close electron systems [1–4], electron–hole systems [5] and normal metal–superconductor systems [6] was measured. In these experiments a current passing through one of the layers (the active layer) induces a current in the other layer (the passive layer), due to the frictional forces. If no current is allowed to flow in the passive layer, a potential bias is developed which cancels the frictional interlayer force. The trans–resistance (per square), or the drag coefficient, is defined as the ratio between the potential developed in the passive layer, to the current in the active layer, i.e. $\rho_D \equiv V_{passive}/I_{active}$. Measurements of $\rho_D$ provide a direct measure of the frictional forces between the two layers. A few mechanisms can lead to frictional drag forces between the charge carriers of different layers. The carriers can interact via Coulomb forces [2,7], phonon exchange [3,8], or magnetic interactions of spontaneously created vortices (in the case of superconducting layers) [9]. The temperature dependence of the observed trans–resistance in electron [2] and electron–hole [5] systems ($\rho_D \propto T^2$) was qualitatively explained by the Coulomb mechanism, although in order to explain the dependence of the trans–resistance on the distance between the layers a virtual phonon drag mechanism was suggested [3,8]. The present paper is limited to the study of the Coulomb drag mechanism only, however, our approach can be, to some extent, generalized to include other mechanisms of drag.

Drag due to the Coulomb interaction between the carriers of spatially separated layers was considered more than a decade ago in Refs. [10,11]. The first direct experimental observation of drag and a comparison with theory was reported in Refs. [1,7]. The new experiments [2,5,6], as well as the exotic behavior of two layer systems in a strong magnetic field [12] have renewed recently the interest in the interlayer friction phenomena.

The previous theoretical studies of the Coulomb drag were based upon the kinetic equation [2,7,13] or upon the Mori memory–function formalism [14]. The latter formulation is sufficiently general to treat also disorder inside the layers. It was found that in dirty sys-tems at a very low temperature, the temperature dependence of the drag coefficient changes from $\rho_D \propto T^2$ to $\rho_D \propto T^2 \ln T$ [14]. In both cases, that of the kinetic equation and that of the memory–function formalism the approximations involved are uncontrollable, i.e. it is not clear what is the order of the corrections to the obtained results. It is also not clear to what extent these approaches can be used to treat quantum interference effects, the presence of an external magnetic field, or intralayer electron correlations. Therefore an alternative rigorous approach is needed. In this paper we employ Kubo linear response formalism in the framework of diagrammatic expansion in order to develop a controlled technique for calculating the drag in various setups. Besides a controlled way to obtain the old results, the diagrammatic approach provides a method to treat the intralayer interactions, quantum interference phenomena *etc*. We obtain a few new results, which can hardly be obtained (if at all) by the other techniques.

The central quantity in the framework of the Kubo linear response is not the trans–resistance, $\rho_D$, but rather the trans–conductance, $\sigma_D$. In order to obtain the experimentally measurable drag coefficient, one should invert the $2 \times 2$ conductance matrix. To lowest non–vanishing order in the interlayer interaction, one obtains

$$\rho_D = \frac{\sigma_D}{\sigma_1 \sigma_2}, \quad (1)$$

where $\sigma_l = (e^2/\hbar) 2 D_l \nu_l$ is the conductance of the $l^{th}$ *isolated* layer ($D$ is the diffusion constant, $\nu$–the density of states per spin). We emphasize the apparent analogy between the drag and the Hall coefficients since both of them are off–diagonal elements of the resistance matrix. As we shall see, this analogy goes very far and is extremely useful. The technical reason is that in both cases one has to calculate "three leg" vertices. The less formal reason is that the dragged current may be considered as a non–dissipative one since the applied voltage and the induced current are spatially separated.

Before sketching the obtained results, let us list the limitations we have restricted ourselves to. We treat the drag effect only to the lowest non–vanishing order in the *interlayer* interactions, while *intralayer* interactions may be included to an infinite order. The interlayer tunneling of electrons is not discussed here. We consider only thin quasi two–dimensional layers (the layer thickness should be less than the screening radius). Generalization to thick layers is not very painful, but requires some effort. We do not consider electron-hole systems, but only two electron layers (possibly with different mobilities and Fermi energies). Only the spatially uniform d.c. drag effect is studied. The drag phenomenon may also occur between two types of carriers which are not spatially separated, e.g., in two band metals [15], heavy and light holes [16] *etc*. Although the diagrammatic approach is certainly useful for studying these examples, our results can not be applied directly.

The previous theoretical studies of the drag phenomenon [7,8,13,14] emphasize the close relation between the drag coefficient and the imaginary part of the intralayer polarization operator (which is related to the dynamic structure factor). It was even proposed [17] that measurements of the drag coefficient may be a useful tool in studying the structure factor of strongly correlated systems. We show, that although the above mentioned relation works surprisingly well (e.g. it reproduces correctly the quantum interference effects and the response to an external magnetic field), it is not generally valid in the presence of intralayer electron–electron correlations. In general, as will be shown, the drag coefficient may be related to the non–linear susceptibilities of the layers. This suggests that the d.c. drag current is a result of rectification by the passive layer of an a.c. fluctuating electric field, created by the active one. In a particular case, when intralayer electron–electron correlations are absent, the non–linear susceptibility is reduced to the product of the diffusion constant and the imaginary part of the polarization operator, reproducing the earlier results.

The role of intralayer correlations is demonstrated for the case in which one or both layers are close to the superconducting transition temperature. This part of our study was motivated by recent experiments on normal metal–superconductor systems [6]. We consider the contribution of spontaneously created fluctuating Cooper pairs to the drag. These lead to a divergent contribution (near the critical temperature) to the conductivity (Aslamazov–Larkin corrections [18]). We have calculated the contribution of these fluctuations to the non–linear susceptibility. It appears to be proportional to a divergent power of $(T - T_c)$ (for dirty two–dimensional films it is $\propto (T - T_c)^{-2}$. This divergence is not strong enough to enhance significantly the drag coefficient in normal metal–superconductor structures and, thus, fails to explain the experiments [6]. We found, however, that in the case of two identical metals, both close to the superconducting transition, the drag coefficient may be enhanced dramatically. The maximal enhancement (within the Ginzburg criterion for validity of the Landau–Ginzburg approach) is by a factor of $\tilde{\sigma}$ – the dimensionless conductance in the normal state.

The outline of the article is as follows. In Sec. II the diagrammatic technique for the drag trans–conductance is developed. The main result of this section is Eq. (6), which establishes the relation between the trans–conductance and the non–linear susceptibility. In Sec. III the non–linear susceptibility, $\vec{\Gamma}$, is calculated for (clean and dirty) normal metals (without intralayer electron–electron interactions). In this case $\vec{\Gamma} \sim D \Im \Pi$ ($\Pi$ is the polarization operator), and the old results are reproduced. The dependence of the trans–resistance on the mobilities, the interlayer distance and the temperature is listed in Sec. III C and summarized in Table I for completeness. In addition to the known results [14], we point out a novel linear temperature dependence of $\rho_D$ in a certain range of temperatures, which takes place in very dirty systems. In Sec. IV we study the quantum interference (weak–localization) corrections to the trans–conductance. Due to a cancellation of a certain class of diagrams, the expression derived in the memory–function formalism [14] appears to be applicable (with the proper renormalization of the diffusion constant). In Sec. V we show that in the absence of intralayer interactions the Hall drag coefficient vanishes (again in agreement with the memory–function formalism). Sec. VI is devoted to the study of fluctuation corrections to the drag coefficient near the superconducting transition. We conclude with a discussion of the results and a brief summary. The appendix is devoted to the calculation of the screened electron–electron interactions in two layer systems. We do not compare the theory with the existing experiments, mainly because it can be found in the literature (see eg. Ref. [4]). The new results obtained here are waiting for an experimental confirmation.

## II. KUBO FORMULATION OF TRANS–CONDUCTANCE

Repeating the standard derivation of the Kubo formula [19] for the case in which an external field is applied to layer 1 and the induced current is measured in layer 2, one obtains for the drag trans–conductance

$$\sigma_D^{ij}(\vec{Q},\Omega) = \frac{1}{\Omega S} \int_0^\infty dt\, e^{i\Omega t} \left\langle \left[ J_1^{i\dagger}(\vec{Q},t), J_2^j(\vec{Q},0) \right] \right\rangle. \quad (2)$$

Here $i,j = \hat{x}, \hat{y}$; $\vec{Q}, \Omega$ are the wave vector and the frequency of the external field, $S$ is the area of the sample and $J_l^i$ is the $i^{th}$ component of the current operator in the $l^{th}$ layer. Diagrams corresponding to Eq. (2) include two separate electron loops with a vector (current) vertex on each one of them, coupled only by the interlayer interaction lines. To second order in the interlayer interaction there are two different diagrams, depicted in Fig. 1. (The first order diagram vanishes when the external field wave vector $\vec{Q}$ is zero. That is the case we are interested in.) We shall focus hereafter on the d.c. trans–conductance $\sigma_D^{ij} \equiv \sigma_D^{ij}(\vec{Q}=0,\Omega \to 0)$.

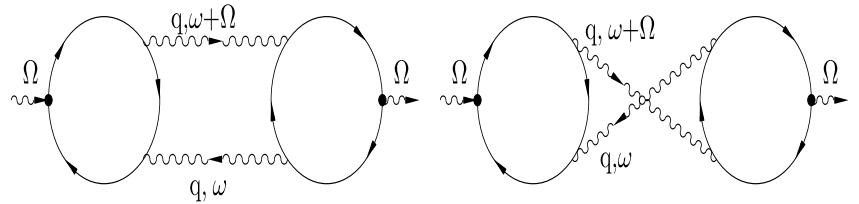

FIG. 1. Two diagrams, contributing to the trans–conductance to second order in the interlayer interaction. The full lines with arrows are the electron Green functions. The wavy lines represent interlayer interactions. Full dots are vector (current) vertices.

Analytically, the two leading order diagrams (Fig. 1) may be written in a symmetric form

$$\sigma_D^{ij} = \frac{T}{2i\Omega_m S} \sum_{\vec{q},\omega_n} \Gamma_1^i(\vec{q},\omega_n+\Omega_m,\omega_n) \Gamma_2^j(\vec{q},\omega_n,\omega_n+\Omega_m) U(\vec{q},\omega_n+\Omega_m) U(\vec{q},\omega_n), \quad (3)$$

where $U(\vec{q},\omega)$ is the interlayer screened Coulomb interaction (see Appendix A) and the vector $\vec{\Gamma}_{1(2)}(\vec{q},\omega_1,\omega_2)$ is the "three leg" object given by the two diagrams depicted in Fig. 2 (the factor 1/2 in Eq. (3) is included in order to prevent double counting).

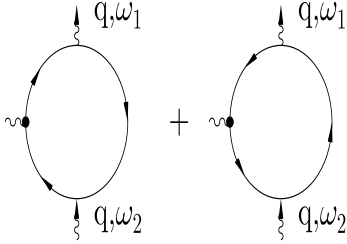

FIG. 2. Diagrams defining the "three leg" vertex, $\vec{\Gamma}(\vec{q},\omega_1,\omega_2) = \vec{\Gamma}(-\vec{q},-\omega_2,-\omega_1)$.

In Eq. (3) the usual Matsubara technique with $\Omega_m = 2\pi i m T$ is employed [20]. After summing over the boson frequencies, $\omega_n$, one should perform an analytical continuation to a real value of $\Omega$, and finally the limit $\Omega \to 0$ should be taken. The sum over $\omega_n$ is done by a contour integration in the complex $\omega$ plane along the contours shown in Fig. 3a. Since the integrals over the small circles cancel the contributions from points $\omega_n = 0$ and $\omega_n = -\Omega_m$ [21], one finds that

$$\sigma_D^{ij} = \frac{1}{2i\Omega_m S}\sum_{\vec{q}} \frac{1}{4\pi i}\oint_c dz \coth\frac{z}{2T}\Gamma_1^i(\vec{q},z+\Omega_m,z)\Gamma_2^j(\vec{q},z,z+\Omega_m)U(\vec{q},z+\Omega_m)U(\vec{q},z)$$

$$= -\frac{1}{8\pi\Omega_m S}\sum_{\vec{q}}\int_{-\infty}^{\infty} d\omega \coth\frac{\omega}{2T}\left[\Gamma_1^{i+-}(\vec{q},\omega,\omega-\Omega_m)\Gamma_2^{j-+}(\vec{q},\omega-\Omega_m,\omega)\right.$$
$$\left. - \Gamma_1^{i+-}(\vec{q},\omega+\Omega_m,\omega)\Gamma_2^{j-+}(\vec{q},\omega,\omega+\Omega_m)\right]U^+(\vec{q},\omega+\Omega_m)U^-(\vec{q},\omega), \quad (4)$$

where the symbols + and − indicate from which part of the complex plane the analytical continuation is performed. The way of analytical continuation of $\vec{\Gamma}^{+-}(\vec{q},\omega_1,\omega_2)$ is illustrated in Fig. 3(b). Note that only the middle part of the contour (cf. Fig. 3(a)) contributes to the Eq. (4). The upper and lower parts of the contour do not contribute, because in the limit $\Omega \to 0$

$$\vec{\Gamma}^{\pm\pm}(\vec{q},\omega,\omega) \propto \frac{\partial}{\partial\vec{q}}\Pi^\pm(\vec{q},\omega) - \frac{\partial}{\partial\vec{q}}\Pi^\pm(\vec{q},\omega) = 0, \quad (5)$$

where $\Pi^\pm(\vec{q},\omega)$ is the retarded (advanced) polarization operator (the two terms in the middle part of the last equation correspond to the two diagrams of Fig. 2). As a result, $\vec{\Gamma}^{\pm\pm}(\vec{q},\omega+\Omega,\omega) \propto \Omega$, and do not contribute to the d. c. trans–conductance.

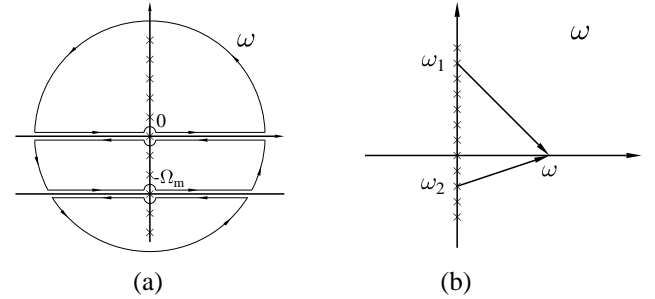

FIG. 3. (a) The contour of integration in the complex $\omega$ plane, employed to perform the sum over the Matsubara boson frequencies in Eq. (3). (b) The way in which the analytical continuation of $\vec{\Gamma}(\vec{q},\omega_1,\omega_2) \to \vec{\Gamma}^{+-}(\vec{q},\omega,\omega)$ is performed.

We perform now the analytical continuation, $\Omega_m \to \Omega$, in Eq. (4) and take the limit $\Omega \to 0$. The result is

$$\sigma_D^{ij} = \frac{1}{16\pi T S}\sum_{\vec{q}}\int_{-\infty}^{\infty}\frac{d\omega}{\sinh^2\frac{\omega}{2T}}\Gamma_1^{i+-}(\vec{q},\omega,\omega)\Gamma_2^{j-+}(\vec{q},\omega,\omega)\left|U^+(\vec{q},\omega)\right|^2. \quad (6)$$

Here we have used the fact that $U^-(\vec{q},\omega) = \left(U^+(\vec{q},\omega)\right)^*$. Applying Onsager relations to $\sigma_D^{ij}(\vec{H})$, one can easily show that

$$\vec{\Gamma}^{+-}(\vec{q},\omega,\omega,\vec{H}) = \vec{\Gamma}^{-+}(\vec{q},\omega,\omega,-\vec{H}), \quad (7)$$

where $\vec{H}$ is an external magnetic field.

Eq. (6) provides the relation between the d. c. trans–conductance, $\sigma_D^{ij}$, and the vertices, $\vec{\Gamma}_{1(2)}^{\pm\mp}(\vec{q},\omega,\omega)$. The latter are the non–linear susceptibilities (namely, rectification) of the electron gases in an external a.c. field. The physical interpretation of this fact is the following. If the passive layer is exposed to an external space and time dependent scalar potential, $\phi(\vec{q},\omega)$, then a d.c. current,

$$\vec{J}_2 = \vec{\Gamma}_2^{-+}(\vec{q},\omega,\omega)|\phi(\vec{q},\omega)|^2, \tag{8}$$

is induced, as a result of the second order rectification effect. In the drag problem the random field, $\phi(\vec{q},\omega)$, originates from the thermal fluctuations of the electron gas in the active layer. In the presence of the external driving field, $\vec{E}_1$, the correlator of the *non-equilibrium* component of the random field is

$$\langle \phi^*\phi \rangle \sim \left(\vec{E}_1 \cdot \vec{\Gamma}_1^{+-}\right) \frac{\partial n_B}{\partial \omega}|U^+|^2, \tag{9}$$

where $n_B(\omega)$ is the Bose distribution function. Substitution of Eq. (9) in Eq. (8) leads to Eq. (6).

The next section is devoted to the calculation of the non-linear susceptibility, $\vec{\Gamma}^{+-}(\vec{q},\omega,\omega)$ in the absence of the intralayer electron-electron interactions and an external magnetic field. It will be shown that, to $O(1/(k_F l))$, $\vec{\Gamma}$ in this case is given by

$$\vec{\Gamma}^{+-}(\vec{q},\omega,\omega) = \vec{\Gamma}^{-+}(\vec{q},\omega,\omega) = e\frac{2D\vec{q}}{\epsilon_F}\Im\Pi^+(\vec{q},\omega), \tag{10}$$

where $D \equiv D(\vec{Q}=0,\Omega=0)$ is the d.c. diffusion constant of the electron gas.

In this particular situation, employing Eqs. (1) and (6), one obtains for the d.c. trans-resistance the result of Refs. [7,13,14]:

$$\rho_D^{ij} = \frac{\hbar}{e^2}\frac{1}{16\pi}\frac{1}{\nu_1\nu_2\epsilon_{F1}\epsilon_{F2}}\frac{1}{TS}\sum_{\vec{q}} q^i q^j \int_{-\infty}^{\infty} \frac{d\omega}{\sinh^2\frac{\omega}{2T}}\Im\Pi_1^+(\vec{q},\omega)\Im\Pi_2^+(\vec{q},\omega)\left|U^+(\vec{q},\omega)\right|^2. \tag{11}$$

We stress, however, that Eq. (10) and hence Eq. (11) are proved below (up to order $O(1/(k_F l))$) only for the case in which electron-electron correlations within each layer are absent. For interacting electrons (e.g. for the drag between two superconductors) one should use the general expression, Eq. (6). The formal reason is that the electron-electron interactions corrections cannot be reduced, in general, to a renormalization of the vertices in the diagrams for $\vec{\Gamma}$, Fig. 2 (see also Fig. 5 bellow).

## III. DRAG BETWEEN NORMAL METALS

We turn now to the calculation of the non-linear susceptibility, $\vec{\Gamma}^{+-}(\vec{q},\omega,\omega)$, for normal metals, without intralayer electron-electron interactions. It will be shown that in this case Eq. (10) may be obtained within controlled approximations.

Recall that $\vec{\Gamma}(\vec{q},\omega_1,\omega_2)$ is given by the two diagrams depicted in Fig. 2 and the analytical continuation should be performed according to Fig. 3(b). The expression corresponding to Fig. 2 has the form

$$\vec{\Gamma}(\vec{q},\omega_1,\omega_2) = T\sum_{\epsilon_n} \text{Tr}\left\{\mathcal{G}_{\epsilon_n}\mathcal{G}_{\epsilon_n+\omega_2}\vec{\tilde{I}}\mathcal{G}_{\epsilon_n+\omega_1} + \mathcal{G}_{\epsilon_n}\mathcal{G}_{\epsilon_n-\omega_1}\vec{\tilde{I}}\mathcal{G}_{\epsilon_n-\omega_2}\right\}, \tag{12}$$

where $\vec{\tilde{I}}$ is the current operator, $\mathcal{G}_{\epsilon_n}$ is the Green function and the trace is taken over the exact quantum states of the disordered system. Next we perform the sum over the fermionic energies, $\epsilon_n$, and the analytical continuation of $\vec{\Gamma}$ according to Fig. 3(b). This is done in a similar way to the one employed for the boson frequency sum in Eq. (3) (cf. also Fig. 3(a)). Then e.g. for the first term on the r.h.s. of Eq. (12) one has

$$\frac{1}{4\pi i}\int d\epsilon \left(\tanh\frac{\epsilon}{2T} - \tanh\frac{\epsilon+\omega_2}{2T}\right) \text{Tr}\left\{\mathcal{G}_\epsilon^+ \mathcal{G}_{\epsilon+\omega_2}^- \vec{\tilde{I}}\mathcal{G}_{\epsilon+\omega_1}^+\right\} \tag{13}$$
$$+ \frac{1}{4\pi i}\int d\epsilon \left(\tanh\frac{\epsilon+\omega_1}{2T} - \tanh\frac{\epsilon}{2T}\right) \text{Tr}\left\{\mathcal{G}_\epsilon^- \mathcal{G}_{\epsilon+\omega_2}^- \vec{\tilde{I}}\mathcal{G}_{\epsilon+\omega_1}^+\right\},$$

where $\mathcal{G}^\pm$ denotes retarded (advanced) single electron Green functions. There are no contributions from $\epsilon_n > -\omega_2 > 0$ and $\epsilon_n < -\omega_1 < 0$ since the trace of three Green functions of the same type vanishes. Performing the analytical continuation according to Fig. 3(b), one obtains

$$\vec{\Gamma}^{+-}(\vec{q},\omega,\omega) = \frac{1}{4\pi i}\int d\epsilon \left(\tanh\frac{\epsilon+\omega}{2T} - \tanh\frac{\epsilon}{2T}\right) \text{Tr}\left[\left(\mathcal{G}_\epsilon^- - \mathcal{G}_\epsilon^+\right)\mathcal{G}_{\epsilon+\omega}^- \vec{\tilde{I}}\mathcal{G}_{\epsilon+\omega}^+\right] \tag{14}$$
$$+ \{\vec{q},\omega \to -\vec{q},-\omega\}.$$

Employing in advance the fact that $\text{Tr}[\cdots]$ does not depend explicitly on $\epsilon$ (for $\epsilon_F\tau \gg 1$), and recalling that in the absence of a magnetic field $\mathcal{G}_\epsilon^- = (\mathcal{G}_\epsilon^+)^*$, one obtains

$$\vec{\Gamma}^{+-}(\vec{q},\omega,\omega) = \frac{\omega}{\pi}\Im\text{Tr}\left[\mathcal{G}_{\epsilon_F}^- \mathcal{G}_{\epsilon_F+\omega}^- \vec{\tilde{I}}\mathcal{G}_{\epsilon_F+\omega}^+\right] + \{\vec{q},\omega \to -\vec{q},-\omega\}. \tag{15}$$

Special care should be taken in the presence of an external magnetic field (see our discussion of the Hall drag coefficient in Sec. V). Further evaluation of the non–linear susceptibility, $\vec{\Gamma}^{+-}(\vec{q}, \omega, \omega)$, should be performed separately for different values of the momentum and energy exchange, $\vec{q}, \omega$.

### A. The diffusive case

To calculate $\vec{\Gamma}^{+-}(\vec{q}, \omega, \omega)$ (see Eq. (15)) in the diffusive regime, i.e., when $\omega \ll 1/\tau$ and $q \ll 1/l$ ($\tau$ and $l = v_F \tau$ are elastic mean free time and path correspondingly), one should employ the impurity diagrammatic technique (see e.g. Ref. [20]). We use here its simplest version with short range impurities only. The extension to the general case is straightforward and we shall comment on it at the end of this section. The dominant diagram is depicted in Fig. 4.

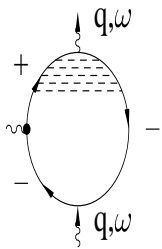

FIG. 4. The leading order (in $(k_F l)^{-1}$) diagram for $\vec{\Gamma}^{+-}(\vec{q}, \omega, \omega)$ in the diffusive regime. The analytical properties of the Green functions are indicated by + (retarded) and − (advanced) signs. Dashed ladder denotes a Diffuson [20].

The corresponding analytical expression is given by

$$\vec{\Gamma}^{+-}(\vec{q}, \omega, \omega) = \frac{2\omega}{\pi} \Im \frac{1}{\tau(Dq^2 - i\omega)} \sum_{\vec{p}} e\vec{v}_p \mathcal{G}_\epsilon^-(\vec{p} - \vec{q}) \mathcal{G}_{\epsilon+\omega}^-(\vec{p}) \mathcal{G}_{\epsilon+\omega}^+(\vec{p}) + \{\vec{q}, \omega \to -\vec{q}, -\omega\} \approx$$
$$\approx \frac{2e}{\pi\tau} \Im \frac{\omega}{Dq^2 - i\omega} \sum_{\vec{p}} \vec{v}_p(-\vec{v}_p \vec{q}) \left(\mathcal{G}_\epsilon^-(\vec{p})\right)^3 \mathcal{G}_\epsilon^+(\vec{p}) + \{\vec{q}, \omega \to -\vec{q}, -\omega\}. \quad (16)$$

In the last line we took an advantage of the fact that $\omega\tau; ql \ll 1$ and expanded up to the first non–vanishing order in these parameters. The momentum sum in the last expression should be performed with some care. The naive evaluation results in: $\sum_{\vec{p}} \ldots = 2\pi S\nu\tau^2 D\vec{q}$ and the two terms on the r.h.s of Eq. (16) cancel each other. The cancellation occurs because the asymmetry between electrons and holes is ignored by the linearization of the spectrum near the Fermi energy. Without such asymmetry the electrons' drag and the holes' drag compensate each other completely. A very similar situation arises in the calculation of the Hall coefficient of dirty metals in the framework of the diagrammatic technique [22,23]. In order to take into account the asymmetry we approximate $\sum_{\vec{p}} \vec{v}_p(-\vec{v}_p \vec{q}) \ldots = -S\nu \vec{q} \int_{-\infty}^{\infty} d\epsilon_p (\epsilon_F + \epsilon_p)/m \ldots$ and obtain $\sum_{\vec{p}} \ldots = 2\pi S\nu\tau^2 D\vec{q}\left(1 + \frac{i}{2\epsilon_F \tau}\right)$. As the leading order does not contribute to $\vec{\Gamma}^{+-}$, one should keep track of terms of the order $(\epsilon_F \tau)^{-1}$. (For a non–parabolic dispersion relation $\epsilon_F$ should be replaced by $(\partial \epsilon_p/\partial p)^2 (2 \partial^2 \epsilon_p/\partial p^2)^{-1}$ at the Fermi energy.) Substituting the result for $\sum_{\vec{p}} \ldots$ into Eq. (16), one obtains

$$\vec{\Gamma}^{+-}(\vec{q}, \omega, \omega) = e\frac{2D\vec{q}}{\epsilon_F} \left[2S\nu \frac{\omega Dq^2}{(Dq^2)^2 + \omega^2}\right], \quad (17)$$

which has precisely the form of Eq. (10), since in the diffusive regime [24]

$$\Pi^+(\vec{q}, \omega) = 2S\nu \frac{Dq^2}{Dq^2 - i\omega}. \quad (18)$$

We conjecture that Eq. (10) is valid if diagrams of the form of Fig. 5(a) are considered. Here (and in the next section) this conjecture is proved for specific examples. The diagrams of the form of Fig. 5(b), which cannot be reduced to a renormalization of the vertices, make Eq. (10) invalid. However, to leading order in $(\epsilon_F \tau)^{-1}$, these diagrams do not contribute. It will be shown in the next section that they do not contribute to weak–localization corrections as well. We conclude thus that Eqs. (10) and (11) are valid up to $O(\frac{1}{\epsilon_F \tau})$ in the diffusive case. One can easily convince himself that the inclusion of long range scattering will not change this conclusion. Indeed, the small angle scattering renormalizes the current vertex and one of the scalar vertices, leading to a redefinition of the diffusion constant only.

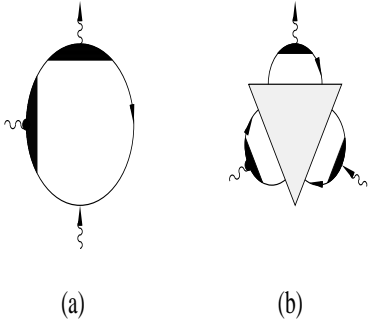

FIG. 5. The general structure of diagrams for the non–linear susceptibility, $\vec{\Gamma}^{+-}(\vec{q},\omega,\omega)$. The black areas represent a vertex renormalization by impurity and intralayer interaction lines. The shaded triangle is an irreducible "three leg" vertex.

In order to perform the calculation of the drag coefficient one should employ the screened interlayer interaction in the diffusive regime which is given by (cf. Appendix A)

$$U^+(\vec{q},\omega) = \frac{1}{S}\frac{\pi e^2 q}{\kappa_1\kappa_2 \sinh qd}\frac{D_1 q^2 - i\omega}{D_1 q^2}\frac{D_2 q^2 - i\omega}{D_2 q^2}, \qquad (19)$$

where the divergences at small $q$ are cut off at $D_{1(2)}q^2 \approx \omega/(\kappa_{1(2)}d)$; here $\kappa_{1(2)}$ are the Thomas–Fermi momenta.

### B. The Ballistic case

Next we consider the ballistic regime where either $q > 1/l$, or $\omega > 1/\tau$. In this case the dominant contribution to $\vec{\Gamma}^{+-}$ comes from diagrams without any internal impurity lines (each Green function is still impurity decorated). Employing the approximation

$$\mathcal{G}^-_{\epsilon+\omega}(\vec{p})\mathcal{G}^+_{\epsilon+\omega}(\vec{p}) = 2\tau\Im\mathcal{G}^-_{\epsilon+\omega}(\vec{p}) \approx 2\pi\tau\delta(\epsilon+\omega-\epsilon_p), \qquad (20)$$

substituting it in Eq. (15) and performing the momentum integration, one obtains for $\omega; q^2/2m \ll \epsilon_F$

$$\vec{\Gamma}(\vec{q},\omega,\omega) = e\frac{2D\vec{q}}{\epsilon_F}\Big[2S\nu\frac{\omega}{v_F q}\theta(v_F q - \omega)\Big], \qquad (21)$$

where $\theta(x)$ is the Heaviside step function. As in the diffusive case the two terms on the r.h.s of Eq. (15) cancel each other in the leading order. To obtain Eq. (21) the non–linearity of the dispersion relation should be taken into account. Note, that Eq. (21) has the form of Eq. (10) again. Indeed, for the ballistic system (at small $q$ and $\omega$) [19]

$$\Im\Pi^+(\vec{q},\omega) = 2S\nu\frac{\omega}{v_F q}\theta(v_F q - \omega). \qquad (22)$$

The screened interlayer interaction in the ballistic case is given by (cf. Appendix A)

$$U^+(\vec{q},\omega) = \frac{1}{S}\frac{\pi e^2 q}{\kappa_1\kappa_2 \sinh qd}. \qquad (23)$$

After establishing the validity of Eq. (10) (and therefore Eq. (11)) for the diffusive and the ballistic cases we shall list results for the drag coefficient for various temperatures and mobilities.

### C. Non–interacting drag trans-resistance for different temperatures and mobilities

In order to understand how the drag coefficient depends on the temperature, the mobilities and the distance between the layers it is instructive to look at the relevant parts of the $(q,\omega)$ plane (see Fig. 6).

We distinguish between two parts of the plane

$$\begin{aligned}\text{Ballistic} &\quad \text{for}\quad \omega < v_F q \quad q > 1/l;\\ \text{Diffusive} &\quad \text{for}\quad \omega < 1/\tau \quad q < 1/l.\end{aligned} \qquad (24)$$

The frequency integration according to Eq. (11) is cut off exponentially at $\omega \approx T$, while the momentum summation is restricted (also exponentially) by $q \approx 1/d$ due to the interlayer interaction potential. We thus distinguish between ballistic, $l \gg d$, and diffusive, $l \ll d$, samples. All the experiments [1,2,5] until the present day were done on ballistic systems.

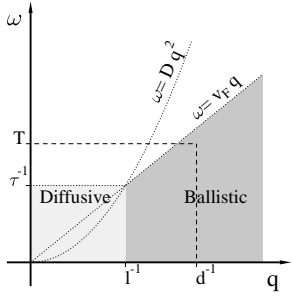

FIG. 6. Domains in the $(q,\omega)$ plane. Dashed area represents (schematically) the regions where $\Im\Pi^+(\vec{q},\omega) \neq 0$.

For ballistic systems ($l \gg d$) the dominant contribution to the trans–resistance, Eq. (11), comes from the ballistic part of the $(q,\omega)$ plane (apart from the exponentially small range of temperatures [14]). Substituting Eqs. (22) and (23) into Eq. (11), one obtains [2,7,13,14]

$$\rho_D = \frac{\hbar}{e^2} \frac{\pi^2 \zeta(3)}{16} \frac{T^2}{\epsilon_{F1}\epsilon_{F2}} \frac{1}{(\kappa_1 d)(\kappa_2 d)} \frac{1}{(k_{F1}d)(k_{F2}d)} . \tag{25}$$

For diffusive systems ($l \ll d$) the entire contribution comes from the diffusive part of the $(q,\omega)$ plane. Substituting Eqs. (18) and (19) into Eq. (11), one obtains for $T \ll \min\{\tau^{-1}, T_0\}$ (where $T_0 \equiv D\min\{\kappa_1,\kappa_2\}/d$ arises due to the divergence of the interaction potential at small momenta, cf. Eq. (19))

$$\rho_D = \frac{\hbar}{e^2} \frac{\pi^2}{12} \frac{T^2}{\epsilon_{F1}\epsilon_{F2}} \ln\frac{T_0}{2T} \frac{1}{(\kappa_1 d)(\kappa_2 d)} \frac{1}{(k_{F1}l_1)(k_{F2}l_2)} . \tag{26}$$

For larger temperatures, $T_0, T \gg \tau^{-1}$, the energy integration is dominated by the region $\omega \approx \tau^{-1}$, the result is

$$\rho_D \approx \frac{\hbar}{e^2} \frac{T\tau^{-1}}{\epsilon_{F1}\epsilon_{F2}} \ln T_0 \tau \frac{1}{(\kappa_1 d)(\kappa_2 d)} \frac{1}{(k_{F1}l_1)(k_{F2}l_2)} . \tag{27}$$

(in order to determine the exact numerical prefactor one should know the behavior of $\Im\Pi^+(\vec{q},\omega)$ for $\omega \approx \tau^{-1}$.) Actually, in this case there is an additional contribution due to two–dimensional plasma modes [25]. In fact, the contribution of the plasmons in this region of temperatures appears to be dominant.

For extremely dirty samples, $l \ll d/\sqrt{\kappa d} \ll d$, a situation where $\tau^{-1} > T \gg T_0$ may occur. The drag coefficient in this case is

$$\rho_D = \frac{\hbar}{e^2} \frac{\pi\zeta(3)}{8} \frac{TT_0}{\epsilon_{F1}\epsilon_{F2}} \frac{1}{(\kappa_1 d)(\kappa_2 d)} \frac{1}{(k_{F1}l_1)(k_{F2}l_2)} . \tag{28}$$

In order to obtain Eq. (28) the full expression for the screened interlayer interaction, Eq. (A2), should be employed.

The dependence of the drag coefficient on the temperature and the interlayer distance is summarized in Table I.

TABLE I. Temperature and interlayer distance dependence of the drag coefficient, $\rho_D$, for different mobilities.

|            | $T \ll \min\{T_0,\tau^{-1}\}$ | $T \gg \min\{T_0,\tau^{-1}\}$ |
|------------|-------------------------------|-------------------------------|
| $l \ll d$  | $d^{-2}T^2 \ln T$             | $d^{-2}T$  [a]                |
| $l \gg d$  | $d^{-4}T^2$  [b]              |                               |

[a] For $T > \tau^{-1}$ the contribution of the plasma modes is dominant [25].
[b] For exponentially small temperatures $T \ll \tau^{-1}\exp\{-3(l/d)^2/4\zeta(3)\}$, $\rho_D \sim d^{-2}T^2 \log T$ [14].

## IV. WEAK–LOCALIZATION CORRECTIONS TO THE DRAG COEFFICIENT

In this section we calculate the weak–localization corrections to the non–linear susceptibility, $\vec{\Gamma}^{+-}(\vec{q},\omega,\omega)$, and to the drag coefficient, $\rho_D$. We restrict ourselves to the diffusion case, $T \ll \tau^{-1}$ and $l \ll d$, as this is the case where the localization corrections can be essential. The drag coefficient, without the localization corrections is given by Eq. (26). Three types of diagram, depicted in Fig. 7(a)–(c), may contribute to the weak–localization corrections to $\vec{\Gamma}^{+-}(\vec{q},\omega,\omega)$.

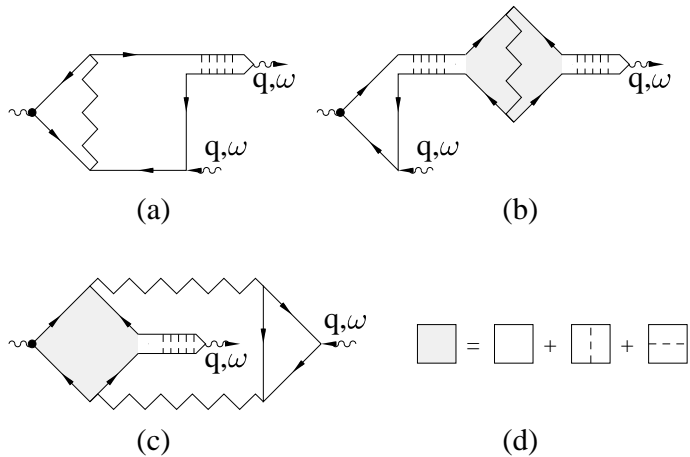

FIG. 7. (a)–(c) The weak–localization corrections to the non–linear susceptibility, $\vec{\Gamma}^{+-}(\vec{q},\omega,\omega)$. The zigzag lines are Cooperons [26], the shaded rectangle (d) is the Hikami box [26,27].

The first two diagrams, Fig. 7(a),(b), belong to the family represented in Fig. 5(a). As we discussed above, such diagrams, lead to the result given by Eq. (10). Indeed a straightforward calculation of diagrams Fig. 7(a),(b) yield

$$\vec{\Gamma}^{+-}(\vec{q},\omega,\omega) = e\frac{2D(0,0)\vec{q}}{\epsilon_F}\Im\left[2S\nu\frac{D(\vec{q},\omega)q^2}{(D(\vec{q},\omega)q^2)^2 + \omega^2}\right], \tag{29}$$

where the renormalized diffusion coefficient $D(\vec{q},\omega)$ for a two–dimensional system is given by [26]

$$D(\vec{q},\omega) = D\left(1 - \frac{1}{\pi k_F l}\ln\frac{1}{\omega\tau}\right). \tag{30}$$

Eq. (29) was verified up to the first order in $\delta D \equiv D(\vec{q},\omega) - D$; one should understand $D(0,0)$ as $D(0,\tau_\phi^{-1})$, where $\tau_\phi$ is a phase relaxation time [23,24].

The diagram in Fig. 7(c), is of the type of Fig. 5(b). It can not be reduced to the product of the d.c. diffusion constant and the polarization operator, like Eq. (10). It is easy to check, however, that due to Hikami–like cancellation [26,27] this diagram does not contain divergent factors like $\ln\tau/\tau_\phi$ or $\ln\omega\tau$. It is not surprising then (this can be checked by a direct calculation [1]) that the diagram in Fig. 7(c) is smaller than the value given by Eq. (29) by a factor of $(k_F l)^{-1}$. We neglect, thus, the contribution of this diagram. As a result, Eq. (29) accounts correctly for the first order weak–localization corrections to the nonlinear susceptibility. This expression has again the form of Eq. (10), ensuring the validity of Eq. (11). Employing Eq. (30), we obtain for the weak–localization corrections to the drag coefficient

$$\frac{\delta\rho_D}{\rho_D} = -\frac{1}{\pi k_{F1} l_1}\ln\frac{1}{2T\tau_1} - \frac{1}{\pi k_{F2} l_2}\ln\frac{1}{2T\tau_2}, \tag{31}$$

where $\rho_D$ is given by Eq. (26). The last equation should be compared with the weak–localization corrections to the conductivity of a two dimensional system [23,26]

$$\frac{\delta\sigma}{\sigma} = -\frac{1}{\pi k_F l}\ln\frac{\tau_\phi}{\tau}. \tag{32}$$

The main difference is that in the case of drag the logarithmic singularity is cut off by $T$ instead of $1/\tau_\phi$. As usual, one expects to have a dependence on a weak magnetic field [23]. In the absence of spin–orbit scattering we predict a *positive* drag magneto–trans–resistance. The characteristic scale of the magnetic field is given by $H_c = T/(eD)$ (in the case of conventional weak–localization it is $H_c = 1/(eD\tau_\phi)$ [23], which is typically much smaller).

---

[1]Technically this calculation is very similar to the calculation of the weak–localization corrections to the Hall coefficient [22,23], where two Cooperon diagrams are also canceled.

The intralayer interaction corrections to the magneto–resistance have typically the same characteristic scale of magnetic fields [24], making the observability of the weak–localization corrections to the drag coefficient problematic.

## V. HALL DRAG

Next we consider the two–layer system in a perpendicular magnetic field. In order to calculate $\sigma_D^{xy}$ we follow the usual treatment of the Hall conductivity in disordered systems [22,23]. The external electric field (in the active layer) and magnetic field (in both layers) are introduced by $\vec{E} = i\Omega \vec{A}_1$ and $\vec{H} = i\left[\vec{k} \times \vec{A}_2\right]$ respectively. The Hall current is proportional to $\vec{E} \times \vec{H} = \Omega\left(\vec{A}_2(\vec{k} \cdot \vec{A}_1) - \vec{k}(\vec{A}_1 \cdot \vec{A}_2)\right)$. It is convenient to choose a gauge where $(\vec{k} \cdot \vec{A}_2) = 0$. The first order corrections ($\propto H$) to the non–linear susceptibility, $\vec{\Gamma}^{+-}(\vec{q},\omega,\omega)$, are given by the three diagrams depicted in Fig. 8. In order to calculate the contributions of these diagrams we expand the Green functions and retain only the contributions which are proportional to $\vec{A}_2(\vec{k} \cdot \vec{A}_1)$. Performing the momentum summation, one should keep track of terms up to order $1/(k_F l)$, since the leading order terms are canceled out. The procedure is similar to the one that was employed in the derivation of Eq. (17).

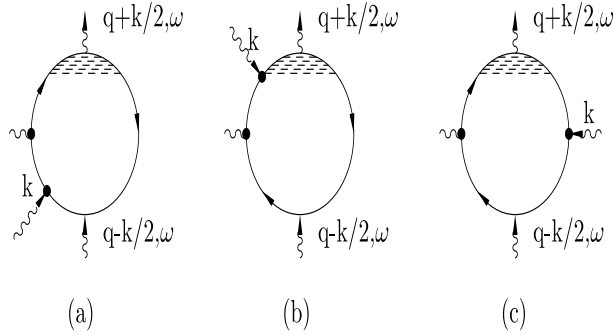

FIG. 8. The three digrams contributing to the first order corrections in the magnetic field to the non–linear susceptibility, $\vec{\Gamma}^{+-}(\vec{q},\omega,\omega)$.

For example, for the diagram Fig. 8(a) one obtains:

$$L_a \sim \sum_{\vec{p}} \vec{v}_p (e\vec{A}_2 \cdot \vec{v}_{p-k/2}) \mathcal{G}^+_{\epsilon+\omega}(\vec{p}) \mathcal{G}^-_\epsilon(\vec{p} - \vec{q} - \vec{k}/2) \mathcal{G}^-_{\epsilon+\omega}(\vec{p} - \vec{k}) \mathcal{G}^-_{\epsilon+\omega}(\vec{p}) \quad (33)$$

$$\approx 2 \sum_{\vec{p}} \vec{v}_p (e\vec{A}_2 \cdot \vec{v}_{p-k/2})(\vec{q} \cdot \vec{v}_p)(\vec{k} \cdot \vec{v}_p) \mathcal{G}^+_\epsilon(\vec{p}) (\mathcal{G}^-_\epsilon(\vec{p}))^5 \to 2e(\vec{A}_2 \cdot \vec{q})\vec{k}\frac{1}{8}S\nu 2\pi\tau^5 v_F\left(1 - \frac{1}{4(\epsilon_F \tau)^2}\right).$$

In the same manner the other diagrams result in $L_b \sim e(\vec{A}_2 \cdot \vec{q})\vec{k}\frac{1}{8}S\nu 2\pi\tau^5 v_F(-2 - \frac{1}{2(\epsilon_F \tau)^2})$, $L_c = 0$. Some care should be taken while performing the above calculations. The point is that in the presence of a magnetic field the contributions from $\mathcal{G}^- \mathcal{G}^- \mathcal{G}^+$ and $\mathcal{G}^+ \mathcal{G}^- \mathcal{G}^+$ (see Eq. (14)) are not complex conjugated and should be calculated separately. (Note that each of these terms separately is not gauge invariant and depends on the particular choice of momentum parameterization (gauge). Only the sum of the two, which is independent of the momentum parameterization is gauge invariant.) Adding all the terms together, one obtains to first order (in the magnetic field) corrections to the non–linear susceptibility

$$\delta\vec{\Gamma}^{+-}(\vec{q},\omega,\omega) = e\frac{2D}{\epsilon_F}\left[\vec{q} \times \vec{H}\right]\frac{e\tau}{m}\Im\Pi^+(\vec{q},\omega). \quad (34)$$

Substituting this in Eq. (6) and employing Onsager relations, Eq. (7), we obtain

$$\sigma_D^{xy} = (\omega_c \tau_1 + \omega_c \tau_2)\sigma_D^{xx}, \quad (35)$$

where $\omega_c = eH/m$ is the cyclotron frequency.

In order to calculate the Hall drag coefficient one has to invert the $4 \times 4$ conductance matrix (with the components $1x, 1y, 2x, 2y$), keeping only the leading order in both the interlayer interaction and the magnetic field. The result is

$$\rho_D^{xy} = 0. \quad (36)$$

The same result can be obtained by direct application of Eq. (11) to the $xy$ component of the trans–resistance [28]. Thus, it has been proven that the off–diagonal part of Eq. (11) is valid as well as the diagonal one (in the absence of the intralayer electron–electron interactions). The fact that $\rho_D^{xy} = 0$ is closely related with the cancellation of the diagram Fig. 8(c) ($L_c = 0$). The physical interpretation of this observation is that the magnetic field does not influence the interlayer interaction but rather affects the "free motion" (diffusion) of the electrons. As a result the Hall drag current is given by a sum of two obvious contributions: the Hall component of the dragged current in the passive layer and the current dragged by the Hall component in the active layer (cf. Eq. (35)). This statement, however, may be not correct for higher orders of interaction corrections.

## VI. DRAG NEAR THE SUPERCONDUCTOR TRANSITION

In this section we consider the Coulomb drag when one (or both) layer is close to (but still above) the transition to the superconducting state. Namely, we shall consider the vicinity of the transition temperature, where the fluctuations of the superconducting order parameter become important. These fluctuations lead to the Aslamazov–Larkin (AL) [18] corrections to the conductivity. For dirty ($1/\tau \ll T_c$) quasi two–dimensional films the AL corrections have a remarkably simple form [18]

$$\delta\sigma_{AL} = \frac{e^2}{2\pi\hbar}\frac{a}{\varepsilon}; \qquad \varepsilon \equiv \frac{T - T_c}{T_c} \ll 1, \qquad (37)$$

where $T_c$ is the transition temperature and $a$ is the phenomenological constant in the Ginzburg–Landau equation (in the framework of BCS theory $a = \pi/8$).

Our aim is to find the fluctuation corrections to the drag coefficient. We restrict ourselves to dirty superconductors, $T_c \ll 1/\tau$. The extension to the clean case is straightforward, but the fluctuation corrections are less pronounced there. As the intralayer (attractive) interactions play a crucial role, one cannot use Eq. (11) and has to employ the general expression, Eq. (6). (Application of Eq. (11) with the AL corrections to the $\Im\Pi^+$ leads to a *wrong* result.) The most divergent (as $T \to T_c$) corrections to the vertex, $\vec{\Gamma}$, are given by the diagram depicted in Fig. 9(a).

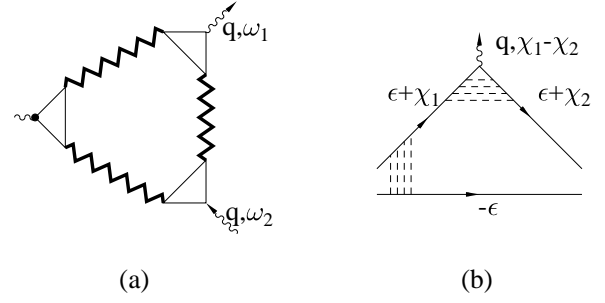

FIG. 9. (a) The fluctuation corrections to $\vec{\Gamma}(\vec{q},\omega_1,\omega_2)$, near the superconducting transition. The bold zigzag lines represent propagators of fluctuating Cooper pairs [18]. (b) The triangular scalar vertex, $B_{\chi_1,\chi_2}$.

This diagram may be considered as the non–linear susceptibility of the Ginzburg–Landau order parameter in an external a.c. field (in a complete analogy with the electron non–linear susceptibility given by the diagram, Fig. 4). The corresponding analytical expression is

$$\delta\vec{\Gamma}(\vec{q},\omega_1,\omega_2) = 8T \sum_{\vec{k};\chi_m} K(\vec{k}-\vec{q},\chi_m) B_{\chi_m,\chi_m+\omega_2} K(\vec{k},\chi_m+\omega_2)\vec{C}_{\vec{k}} K(\vec{k},\chi_m+\omega_1) B_{\chi_m+\omega_1,\chi_m}$$
$$+ \{\vec{q},\omega_1,\omega_2 \to -\vec{q},-\omega_1,-\omega_2\}, \qquad (38)$$

where the factor 8 is a result of the charge of a Cooper pair and the spin summation. In the above equation $K(\vec{k},\chi_m)$ is a propagator of the fluctuating Cooper pairs,

$$\vec{C}_{\vec{k}} = eS\nu\frac{a}{T_c}2D\vec{k} \qquad (39)$$

is the vector triangular vertex [18] (see Fig. 9(a)) and $B_{\chi_1,\chi_2}$ is the scalar triangular vertex, Fig. 9(b). Integrating over a contour with three cut lines, $\chi = 0, -\omega_1, -\omega_2$, in the complex $\chi$ plane and performing analytical continuation according to Fig. 3(b), we obtain

$$\delta\vec{\Gamma}^{+-}(\vec{q},\omega,\omega) = 8\sum_{\vec{k}} \frac{\vec{C}_{\vec{k}}}{4\pi i} \int_{-\infty}^{\infty} d\chi \left(\coth\frac{\chi+\omega}{2T} - \coth\frac{\chi}{2T}\right) K^-(\vec{k},\chi)K^+(\vec{k},\chi) \qquad (40)$$

$$\left[K^+(\vec{k}-\vec{q},\chi+\omega)B^{++}_{\chi,\chi+\omega}B^{+-}_{\chi+\omega,\chi} - K^-(\vec{k}-\vec{q},\chi+\omega)B^{+-}_{\chi,\chi+\omega}B^{--}_{\chi+\omega,\chi}\right] + \{\vec{q},\omega \to -\vec{q},-\omega\}.$$

The terms with $K^+K^+K^+$ and $K^-K^-K^-$ cancel the contributions of the unphysical poles at $\chi = 0$ and $\chi = -\omega$ out. In the close vicinity of the transition, $\varepsilon \ll 1$, one may approximate

$$\coth\frac{\chi+\omega}{2T} - \coth\frac{\chi}{2T} \approx -\frac{2T_c\omega}{\chi(\chi+\omega)}. \qquad (41)$$

If we perform now the integration in Eq. (40), employing the standard expressions for the fluctuation propagators, $K^\pm$, [18], we shall obtain an exact cancellation, i.e., $\vec{\Gamma}^{+-} = 0$ (after the term with $\vec{q}, \omega \to -\vec{q}, -\omega$ is taken into account). This is again due to linearization of the spectrum (see Sec. III A). The same problem arises in the calculation of the fluctuation corrections to the Hall coefficient [29–31]. It was shown [31], based only on the gauge invariance of the Ginzburg–Landau equation, that the fluctuation propagator, which takes into account the electron–hole asymmetry, has the form

$$K^\pm(\vec{k},\chi) = -\frac{1}{S\nu}\left[\frac{a}{T_c}Dk^2 \mp i\chi\left(\frac{a}{T_c} \mp \frac{i}{2}\frac{\partial \ln T_c}{\partial \mu}\right) + \varepsilon\right]^{-1}, \qquad (42)$$

where $\mu$ is a chemical potential. Note, that $\vec{C}_{\vec{k}} = -e\partial K^{-1}/\partial \vec{k}$. The form of the propagator, Eq. (42), suggests that the scalar vertex, $B \propto \partial K^{-1}/\partial \chi$, should also contain the asymmetry factor, $\partial \ln T_c/\partial \mu$, [32]. An explicit calculation of the triangular scalar vertex, Fig. 9(b), leads to the following results

$$B^{\pm\pm}_{\chi_1,\chi_2} = \mp iS\nu\left[\frac{a}{T_c}\frac{-i(\chi_1-\chi_2)}{Dq^2-i(\chi_1-\chi_2)} \mp \frac{i}{2}\frac{\partial \ln T_c}{\partial \mu}\right];$$

$$B^{\pm\mp}_{\chi_1,\chi_2} = \mp iS\nu\left[\frac{a}{T_c}\frac{-i(\chi_1+\chi_2)}{Dq^2-i(\chi_1-\chi_2)} \mp \frac{i}{2}\frac{\partial \ln T_c}{\partial \mu}\right]. \qquad (43)$$

Next we substitute Eqs. (39),(41)–(43) into Eq. (40) and perform the energy integration. Leaving only the most divergent (as $\varepsilon \to 0$) terms, we obtain to first order in the asymmetry factor, $\partial \ln T_c/\partial \mu$,

$$\delta\vec{\Gamma}^{+-}(\vec{q},\omega,\omega) = e\,2D\vec{q}\frac{8a^2}{T_c}\frac{\partial \ln T_c}{\partial \mu}\frac{\omega Dq^2}{(Dq^2)^2+\omega^2}\sum_{\vec{k}}\frac{1}{\Upsilon_+\Upsilon_-}\frac{\Upsilon_+ + \Upsilon_-}{(\Upsilon_+ + \Upsilon_-)^2+(a\omega/T_c)^2}, \qquad (44)$$

where

$$\Upsilon_\pm \equiv \frac{a}{T_c}D(\vec{k}\pm\frac{\vec{q}}{2})^2 + \varepsilon.$$

We shall see that the dominant contribution to the drag trans–conductance comes from an extremely small interlayer momentum exchange, $Dq^2 \approx (T-T_c)/(\kappa d) \ll (T-T_c)$. The intralayer momentum exchange, $\vec{k}$, has a characteristic value given by $Dk^2 \approx \omega \approx (T-T_c)$. Since $q \ll k$ the summation over $\vec{k}$ in Eq. (44) may be easily performed, leading (for the two dimensional case) to

$$\delta\vec{\Gamma}^{+-}(\vec{q},\omega,\omega) = e\frac{2D\vec{q}}{\left(\frac{\partial \ln T_c}{\partial \mu}\right)^{-1}}\left[2S\nu\frac{\omega Dq^2}{(Dq^2)^2+\omega^2}\right]\phi\left(\frac{a\omega}{2T_c\varepsilon}\right)\frac{a}{\tilde\sigma_s\varepsilon^2}, \qquad (45)$$

where

$$\phi(x) = \frac{1}{2\pi x^2}\ln(1+x^2)$$

and $\tilde\sigma_s = 2D\nu$ is the dimensionless conductance of a film in a normal state. Eq. (45) (cf. with Eq. (17)) solves the problem of finding the fluctuation corrections to the non–linear susceptibility (for $Dq^2 \ll \omega$) in dirty superconductors. In the next subsections we shall incorporate it with the general formula, Eq. (6), in order to find the drag trans–conductance between a normal metal and a superconductor and between two identical superconductors. Let us stress that Eq. (45) can not be obtained by a substitution of the fluctuation corrections to $D(\vec{q},\omega)$ into Eq. (16). The latter procedure leads to a wrong conclusion, $\delta\vec{\Gamma}^{+-} \propto \varepsilon^{-1}$, instead of $\delta\vec{\Gamma}^{+-} \propto \varepsilon^{-2}$.

The validity of the perturbation theory is restricted by the condition $\delta\sigma_{AL} \ll \sigma_s$, i.e. $\varepsilon \gg Gi$, where $Gi = a/(2\pi\tilde\sigma_s)$ is the Ginzburg parameter for dirty two–dimensional superconductors [30]. In the temperature range of

$$Gi \ll \varepsilon \ll \sqrt{Gi} \qquad (46)$$

the fluctuation corrections to the non–linear susceptibility become dominant, i.e. $\delta\vec{\Gamma} \gg \vec{\Gamma}$ (see Ref. [30] for a discussion of the analogous situation in the case of the Hall coefficient).

### A. Superconductor–Normal Metal Drag

As we already mentioned above, the fluctuation corrections to the drag coefficient come from extremely small energy and momentum interlayer exchange, $\omega \approx (T-T_c) \ll 1/\tau$;

$Dq^2 \approx (T - T_c)/(\kappa d) \ll 1/\tau$. Because of this one should employ expressions derived for the diffusive regime, Sec. III A, irrespective of the ratio between $d$ and $l$. Substituting Eqs. (17), (45) and (19) into Eq. (6), we obtain the fluctuation corrections to the drag trans–conductance

$$\delta\sigma_D^{s-n} = \frac{e^2}{\hbar} \frac{1}{8\pi^2} \frac{T_c^2}{\epsilon_{Fn} \left(\frac{\partial \ln T_c}{\partial \mu}\right)^{-1}} \frac{\ln(\kappa d)}{(\kappa_n d)(\kappa_s d)} \frac{1}{\tilde{\sigma}_s \varepsilon}, \qquad (47)$$

where the subscripts $s$ and $n$ refer to the superconductor and the normal metal respectively; $\kappa \equiv \min\{\kappa_n; \kappa_s\}$. Comparison with Eqs (26) and (37) shows that up to logarithmical multipliers

$$\frac{\delta\sigma_D^{s-n}}{\sigma_D} \approx \frac{\delta\sigma_{AL}}{\sigma_s} \approx \frac{1}{\tilde{\sigma}_s \varepsilon}. \qquad (48)$$

Since the corrections to the drag coefficient are given by

$$\delta\rho_D^{s-n} = \frac{\delta\sigma_D^{s-n}}{\sigma_n(\sigma_s + \delta\sigma_{AL})}, \qquad (49)$$

they are at most of the same order of magnitude as $\rho_D$ in the normal state. Note that, due to the integration over the full $(\vec{q}, \omega)$ phase space, $\delta\rho_D^{s-n} < \rho_D$ even if $\delta\vec{\Gamma} \gg \vec{\Gamma}$. Hence, we think that the fluctuation corrections to the Coulomb drag coefficient cannot explain the recent experiments [6] (for an alternative explanation see [9]).

### B. Superconductor–Superconductor Drag

Next we consider the drag effect between two identical superconductors. We confine ourself to the temperature region $Gi \ll \varepsilon \ll \sqrt{Gi}$, where $\delta\vec{\Gamma} \gg \vec{\Gamma}$. Substituting Eqs. (45) and (19) into Eq. (6) and performing the integrations, we obtain

$$\delta\sigma_D^{s-s} = \frac{e^2}{\hbar} \frac{1 - \ln 2}{12\pi^3} \frac{T_c^2}{\left(\frac{\partial \ln T_c}{\partial \mu}\right)^{-2}} \frac{\ln(\kappa_s d)}{(\kappa_s d)^2} \frac{a}{\tilde{\sigma}_s^2 \varepsilon^3}. \qquad (50)$$

In the present case the divergence appears to be extremely pronounced, $\propto \varepsilon^{-3}$. As a result the fluctuation corrections to the drag coefficient may exceed essentially it's normal state value. For $\varepsilon \approx Gi$ the enhancement is $\delta\rho_D^{s-s}/\rho_D \approx \tilde{\sigma}_s \gg 1$. The normal state conductance, $\tilde{\sigma}_s$, cannot be arbitrarily large since the dirty film condition, $T_c \ll 1/\tau$, should be fulfilled.

## VII. DISCUSSION OF THE RESULTS

In this section we summarize and discuss the main results of this paper. First, the regular diagrammatic treatment of the interlayer Coulomb drag to the lowest non–vanishing order in the interlayer interaction has been developed. The method that was developed may be useful in other situations where two kinds of mutually interacting carriers are important (e.g. two band materials). Non–Coulomb mechanisms of drag may be easily treated within the same scheme (to second order in the interlayer interaction).

It has been shown that the calculation of the drag trans–conductance may be reduced to the calculation of the non–linear susceptibility (rectification) of the electron gas, Eq. (6). This fact reflects the physical mechanism of the drag phenomenon. Namely, the d.c. drag current is a result of the rectification by the passive layer of the fluctuating electric field, created by the active one. In special cases, the non–linear susceptibility may be reduced to the product of the diffusion constant and the polarization operator, Eq. (10), which reproduces the earlier result, Eq. (11), derived using the kinetic equation [7,13], and the memory function formalism [14]. We conjecture that this reduction is valid, when the correlations between the electrons inside each layer may be neglected. The above conjecture was checked by a direct calculation of the non–linear susceptibility for diffusive and ballistic systems. It proved to be valid even for the off–diagonal trans–conductance (Hall drag) and the weak–localization corrections. In the last two cases the validity of Eq. (10) is not obvious *a priori* and reveals itself through a subtle cancellation of certain classes of diagrams (Figs. 7(c) and 8(c)). For completeness the dependence of $\rho_D$ on the temperature, the mobilities and the interlayer distance in the absence of intralayer interactions are listed in Sec. III C and summarized in Table I. As a byproduct we point out the novel linear temperature dependence of the drag coefficient, taking place in extremely dirty samples, $l < \sqrt{d/\kappa}$.

Next a system where the intralayer electron–electron correlations are of crucial importance is studied. As an example, metals just above the transition to the superconducting state were investigated. Fluctuations of the superconductor order parameter cause cor-rections to the non–linear susceptibility. The dragged current may be enhanced due to spontaneously created unstable Cooper pairs. The non–linear susceptibility is found to diverge as $(T-T_c)^{-2}$ close to the superconductor phase transition. Note that Eq. (10) which works perfectly well for the non–interacting case, gives a wrong result, $\propto (T-T_c)^{-1}$. Having the result for the susceptibility, one can easily calculate the fluctuation corrections to the drag trans–conductance between a normal metal and a superconductor and between two superconductors. In the first case the correction to the drag trans–conductance appears to be proportional to $(T-T_c)^{-1}$. As a result the drag trans–resistance (Eq. (49)) is not enhanced significantly. Apparently the fluctuation corrections can not explain the recent experiments [6] (for an alternative explanation see Ref. [9]). In the case of two superconductors, however, the divergence of the trans–conductance appears to be much stronger, $\propto (T-T_c)^{-3}$. As a result the trans–resistance may be enhanced by a factor of $\check{\sigma}_s \gg 1$ ($\check{\sigma}_s$ is the normal state dimensionless conductance of the film). We believe that the last prediction may be checked experimentally. The example of superconductors shows that the relation between the drag coefficient and the polarization operators, Eq. (11), may be incorrect for strongly correlated systems.

## VIII. ACKNOWLEDGMENTS

Discussions with E. Shimshoni and D. E. Khmelnitskii have initiated this project. We are especially grateful to A. G. Aronov for his deep insight and useful suggestions. Discussions with A. Entelis, A. M. Finkel'stein, Y. Gefen, A. I. Larkin, Y. Levinson, A. H. MacDonald, M. Yu. Reizer, M. Rokni and U. Sivan are highly acknowledged. This research was supported by the German–Israel Foundation (GIF) and the U.S.–Israel Binational Science Foundation (BSF).

## APPENDIX A: SCREENED INTERLAYER INTERACTION

In this Appendix we treat the screened Coulomb interaction in two layer systems. To this end we employ the standard random phase approximation (RPA) [19].

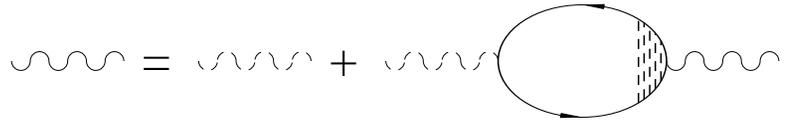

FIG. 10. Screened Coulomb interactions in the RPA. The dashed and full wavy lines are the bare and the screened interactions correspondingly.

The Dyson equation (see Fig. 10) in the RPA is most conveniently written in a matrix form

$$\begin{pmatrix} V_1 & U \\ U & V_2 \end{pmatrix} = \begin{pmatrix} V^0 & U^0 \\ U^0 & V^0 \end{pmatrix} - \begin{pmatrix} V^0 & U^0 \\ U^0 & V^0 \end{pmatrix} \begin{pmatrix} \Pi_1 & 0 \\ 0 & \Pi_2 \end{pmatrix} \begin{pmatrix} V_1 & U \\ U & V_2 \end{pmatrix}. \quad (A1)$$

Here $V^0 = 2\pi e^2/(Sq)$ and $U^0 = V^0 e^{-qd}$ are the bare intralayer and interlayer interactions correspondingly; $V_l$ and $U$ are the screened intralayer and interlayer interactions, $\Pi_l(\vec{q},\omega)$ is polarization operator of the $l^{th}$ layer. Since the tunneling between the layers is neglected the off diagonal elements of the $\Pi$ matrix are equal to zero. Solving the RPA equation and employing the bare interaction value, we obtain for the screened interlayer interaction

$$U^+(\vec{q},\omega) = \left[ 2\Pi_1^+(q,\omega)\Pi_2^+(q,\omega)\sinh(qd)\frac{2\pi e^2}{Sq} + \left(\frac{Sq}{2\pi e^2} + \Pi_1^+(q,\omega) + \Pi_2^+(q,\omega)\right) e^{qd} \right]^{-1}. \quad (A2)$$

In order to obtain Eqs. (19) and (23) we substitute $\Pi^+(q,\omega) = 2S\nu Dq^2/(Dq^2 - i\omega)$ for the diffusive system, or $\Pi^+(q,\omega) = 2S\nu$ for the ballistic one and retain the leading order in $\kappa d \gg 1$ ($\kappa \equiv 4\pi e^2 \nu$ is the inverse Thomas–Fermi screening radius).